\documentclass[conference]{IEEEtran}
\IEEEoverridecommandlockouts
\usepackage{cite}
\usepackage{amsmath,amssymb,amsfonts}
\usepackage{algorithmic}
\usepackage{graphicx}
\usepackage{textcomp}
\usepackage{xcolor}
\usepackage{multirow}

\def\BibTeX{{\rm B\kern-.05em{\sc i\kern-.025em b}\kern-.08em
    T\kern-.1667em\lower.7ex\hbox{E}\kern-.125emX}}

    \title{Synthetic generation of 2D data records based on Autoencoders*\\
    
    \thanks{\footnotesize \textsuperscript{*}TeChBioT project n°101103176. Funded by the European Union. Views and opinions expressed are however those of the author(s) only and do not necessarily reflect those of the European Union or the European Defence Fund. Neither the European Union nor the granting authority can be held responsible for them.}
    
}
\author{\IEEEauthorblockN{1\textsuperscript{st} Darius Couchard}
\IEEEauthorblockA{\textit{Royal Military Academy (RMA)}\\
Brussels, Belgium \\
darius.couchard@mil.be}
\and
\IEEEauthorblockN{2\textsuperscript{nd} Oscar Olarte}
\IEEEauthorblockA{\textit{Royal Military Academy (RMA)}\\
Brussels, Belgium \\
oscar.olarterodriguez@mil.be}
\and
\IEEEauthorblockN{3\textsuperscript{rd} Rob Haelterman}
\IEEEauthorblockA{\textit{Royal Military Academy (RMA)}\\
Brussels, Belgium \\
rob.haelterman@mil.be}
}

\newcommand{\preprintdisclaimer}{%
  \vspace*{-\baselineskip} % Adjust spacing if needed
  \noindent\textit{This work has been submitted to the IEEE for possible publication. Copyright may be transferred without notice, after which this version may no longer be accessible.}\\[\baselineskip]
}

\begin{document}

\preprintdisclaimer

\maketitle

\begin{abstract}
    Gas Chromatography coupled with Ion Mobility Spectrometry (GC-IMS) is a dual-separation analytical technique widely used for identifying components in gaseous samples by separating and analysing the arrival times of their constituent species. Data generated by GC-IMS is typically represented as two-dimensional spectra, providing rich information but posing challenges for data-driven analysis due to limited labelled datasets. This study introduces a novel method for generating synthetic 2D spectra using a deep learning framework based on Autoencoders. Although applied here to GC-IMS data, the approach is broadly applicable to any two-dimensional spectral measurements where labelled data are scarce. While performing component classification over a labelled dataset of GC-IMS records, the addition of synthesized records significantly has improved the classification performance, demonstrating the method's potential for overcoming dataset limitations in machine learning frameworks.\\
\end{abstract}


\section{Introduction}

Gas Chromatography coupled with Ion Mobility Spectrometry (GC-IMS) is a technique
used to identify chemical components within a sample \cite{bibgcims}. Initially, the sample, carried by
a carrier gas, is introduced into the GC column, where interactions between the sample
components and the column affect their transit speeds, leading to an initial separation.
Upon exiting the GC column, the sample enters the IMS device, where particles are ionized and accelerated through an electric field, achieving a further level of separation \cite{bibims}.

\begin{figure}[htbp]
    \centerline{\includegraphics[width=0.45\textwidth]{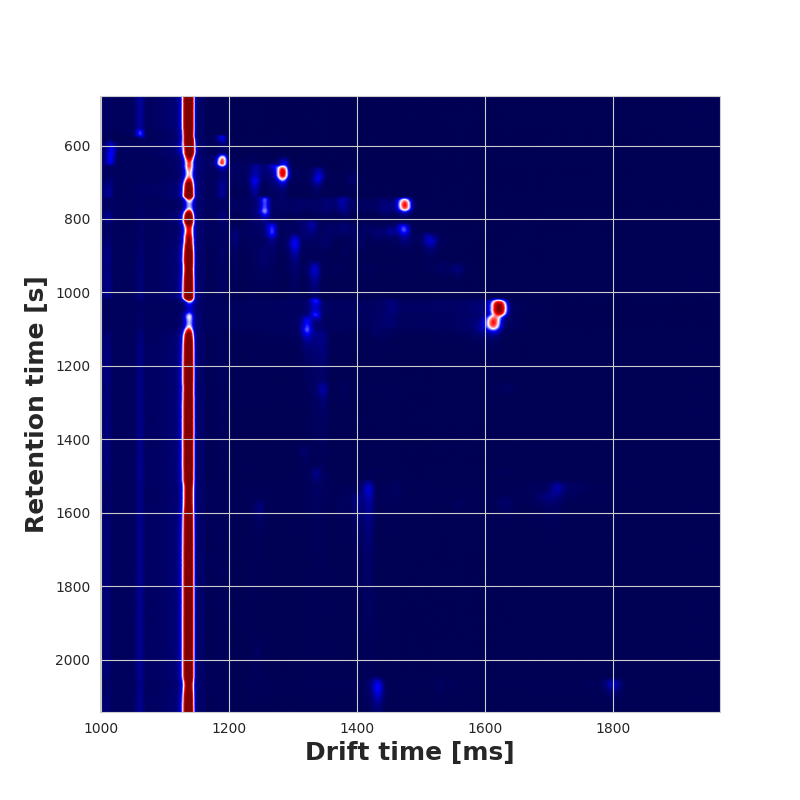}}
    \caption{Section of a GC-IMS Record containing Lactobacillus Brevis. The location of the intensity peaks is the signature of the organism over the 2D spectrum.}
    \label{gcims}
\end{figure}

The GC-IMS device generates data in the form of a 2D matrix, as illustrated in Fig. \ref{gcims}, where one axis represents
the separation achieved by the GC column, referred as the retention time, and the
other axis corresponds to the separation provided by the IMS device, known as the drift
time.

Nontargeted analysis is a method which utilizes the entire matrix to classify a record. The nontargeted approach can be implemented by machine learning pipelines, enabling the creation of a classification model directly from data rather than relying on an expert-driven system. However, available datasets are limited, and the dimensionality of GC-IMS matrices is notably high.

State-of-the-art methods, such as diffusion models, have demonstrated effectiveness in augmenting supervised datasets for traditional RGB images \cite{bibdiffusion}. However, these models typically require extensive datasets for training. Moreover, models designed for traditional RGB images are often not transferable to 2D spectral data due to the fundamental differences in image characteristics, therefore making transfer learning difficult to generalize to this type of records. Although simulation software exists for certain measurement fields, such as SimFCS for fluorescence correlation spectroscopy \cite{b5}, many measurement techniques lack similar tools. For instance, no simulation software has been developed to date for GC-IMS. Creating new simulation software requires substantial resources and specialized expertise, which are often unavailable. This highlights the need for data-efficient, field-agnostic methodologies specifically tailored for the generation of 2D spectra. This work introduces a novel approach for record synthesis using a double autoencoder architecture coupled with a latent matrix resampling method. We demonstrate that incorporating these synthetic records into a classification pipeline improves performance by increasing dataset variability and enhancing the robustness of the resulting classification model.

%In the present work, a double-autoencoder architecture and a latent matrix resampling method for record synthetisation are introduced. It is then demonstrated that incorporating the novel records into a classification pipeline enhances performance by increasing dataset variability, thereby making the classification model more robust.

\section{Methodology}

Synthesizing realistic GC-IMS records involves preserving common features shared across all records and classes, while including variable features, such as the position of the peaks. To achieve this, it is helpful to reduce data to a lower dimensionality, as the original dimensionality of 2D spectra can be high and contain redundant information across records. Methods such as Principal Component Analysis (PCA) can reduce the dimensionality from a linear projection. However, since the features in the 2D spectrum are not independent, employing a non-linear approach like autoencoders can capture more intricate patterns \cite{bibautoencoder}.

As the final objective is to obtain synthesized 2D spectra, the process of dimensionality reduction must be reversible. This constitutes an additional rationale to use an autoencoder architecture, as the records can be reconstructed, albeit with a loss, from the reduced dimensionality.

\subsection{Autoencoders}

\begin{figure}[htbp]
    \centerline{\includegraphics[width=0.25\textwidth]{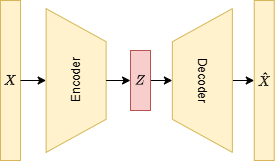}}
    \caption{Autoencoder architecture consisting of an Encoder, a Decoder, and a vector \(\mathbf{z}\) representing the \textit{bottleneck}. The training process ensures that \(\mathbf{z}\) encodes nonredundant information while maintaining a limited data volume, by optimizing the model to minimize the Mean Squared Error between the reconstructed signal \(\mathbf{\hat{X}}\) and the original signal \(\mathbf{X}\).}
    \label{autoencoder}
\end{figure}

Autoencoders are a type of deep learning architecture designed for unsupervised learning. They are primarily used for dimensionality reduction and feature extraction. The architecture is composed of two main components: a parametrized \textit{encoder} $f_\theta$ and a parametrized \textit{decoder} $g_\phi$. The encoder maps input data $\mathbf{x} \in \mathbb{R}^l$ to a lower-dimensional representation named \textit{latent space}, also named \textit{bottleneck} $\mathbf{z} = f_\theta(\mathbf{x})$ where $\mathbf{z} \in \mathbb{R}^d$, and $d < l$. The decoder then reconstructs the input from the latent representation: $\mathbf{\hat{x}} = g_\phi(\mathbf{z})$ where $\mathbf{\hat{x}} \in \mathbb{R}^l$. (Fig. \ref{autoencoder}).

The parameters $\theta$ and $\phi$ of the encoder and decoder functions respectively are learned during the training process. This involves fitting the model to the available dataset by adjusting these parameters to minimize the Mean Squared Error (MSE) along all data points $n$:

\begin{equation}
    \mathrm{MSE}(\theta, \phi) = \frac{1}{n} \sum_{i=1}^n \| \mathbf{x}_i - \mathbf{\hat{x}}_i \|^2 = \frac{1}{n} \sum_{i=1}^n \| \mathbf{x}_i - g_{\phi}(f_{\theta}(\mathbf{x}_i)) \|^2
    \label{mse}
\end{equation}

There are many architectures of autoencoders available in the literature; they vary from the input data that they support to the actual neural network implementation \cite{bibdenoising}, \cite{bibvariational}.

\subsection{Sequential Autoencoders}

Autoencoders designed for image analysis, such as Convolutional Neural Network (CNN) \cite{bibcnnautoencoder}, typically require large datasets to converge effectively while avoiding overfitting. However, GC-IMS measurements contain strong correlations within the rows of the matrix, as these rows represent the drift time, corresponding to the IMS component of the device.

% The first autoencoder encodes each row into latent vectors, and the second encodes the evolution of each value in the latent vectors in the direction of the columns, yielding a \textit{latent matrix}.

Therefore, an architecture composed of two sequential autoencoders is employed. Specifically, for each spectrum represented by a matrix $\mathbf{X}^{(i)} \in \mathbb{R}^{m \times n}$, where $i$ is the index of the matrix in the dataset, columns are extracted as timeseries and individually encoded into latent space vectors of size $d$ where $d < m$ by using the first autoencoder $f_\theta : \mathbb{R}^m \rightarrow \mathbb{R}^d$. This results in an intermediate matrix $\mathbf{Z}^{(i)} \in \mathbb{R}^{d \times n}$.

\begin{figure}[htbp]
    \centerline{\includegraphics[width=0.45\textwidth]{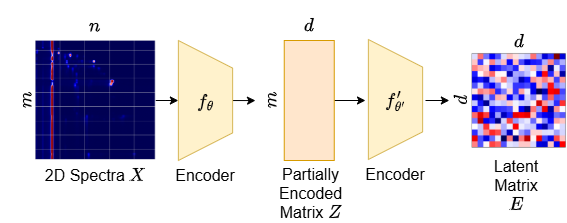}}
    \caption{Use of encoders $f_\theta$ and $f'_{\theta'}$ to obtain the latent matrix. The columns of the spectra $\mathbf{X}^{(i)}$ are individually encoded with $f_\theta: \mathbb{R}^m \rightarrow \mathbb{R}^d$. The rows of the resulting matrix are then encoded with $f'_{\theta'}: \mathbb{R}^n \rightarrow \mathbb{R}^d$. The final latent matrix $\mathbf{E}^{(i)} \in \mathbb{R}^{d \times d}$ is a compressed representation of the original 2D spectra. Two sequential autoencoders are trained on timeseries derived from the records instead of one autoencoder on the entire image, providing more training data and exploiting correlations within IMS scans.}
    \label{seq-ae}
\end{figure}

Subsequently, a second autoencoder is applied to the rows of $\mathbf{Z}^{(i)}$, capturing the evolution of each latent space value computed by the first autoencoder. Using a second encoder $f'_{\theta'} : \mathbb{R}^n \rightarrow \mathbb{R}^k$, where $k < n$ the rows of the partially encoded matrix are also extracted and individually transformed, yielding a fully encoded \textit{latent matrix} $\mathbf{E}^{(i)} \in \mathbb{R}^{d \times k}$. For simplification purposes in the implementation of the pipeline, the latent vector size of the second autoencoder is set to be the same as that of the first autoencoder, i.e., $k = d$. Fig. \ref{seq-ae} displays how both autoencoders are sequentially employed.

Finally, from the latent matrix, an approximation of the original 2D spectra $\mathbf{\hat{X}} \in \mathbb{R}^{m \times n}$ can be reconstructed by sequentially employing the decoders $g'_{\phi'}$ and $g_\phi$, which are trained simultaneously with the encoders $f'_{\theta'}$ and $f_\theta$.

\begin{equation}
    \mathbf{\hat{X}^{(i)}} =  g_\phi \circ g'_{\phi'}(\mathbf{E^{(i)}})
    \label{reconstruct-eq}
\end{equation}

% Essentially, each spectra $\mathbf{X}^{(i)} \in \mathbb{R}^{m \times n}$ where $m$ is the number of rows and $n$ the number columns, is treated as a set $\mathcal{X}^{(i)} = [ x^{(i)}_{1,:}, x^{(i)}_{2,:}, \dots, x^{(i)}_{m,:} ]$ on which each element $x^{(i)}_{l, :}$ is the $l$-th row of matrix $\mathbf{X}^{(i)}$.

\subsection{Synthesization}

Once all records from the available dataset have been encoded, latent matrices \(\mathbf{E}^{(i)}\) are grouped according to their label, which is the chemical composition of their corresponding gas sample. Let \(\mathcal{L}\) represent the set of all existing labels. The set of records corresponding to a particular label \(\ell \in \mathcal{L}\) is denoted as:

\begin{equation}
    \mathbf{E}_\ell = \{\mathbf{E}^{(i)} \mid \text{label of } \mathbf{X}^{(i)} = \ell \}, \quad \ell \in \mathcal{L}.
\end{equation}

\begin{figure}[htbp]
    \centerline{\includegraphics[width=0.5\textwidth]{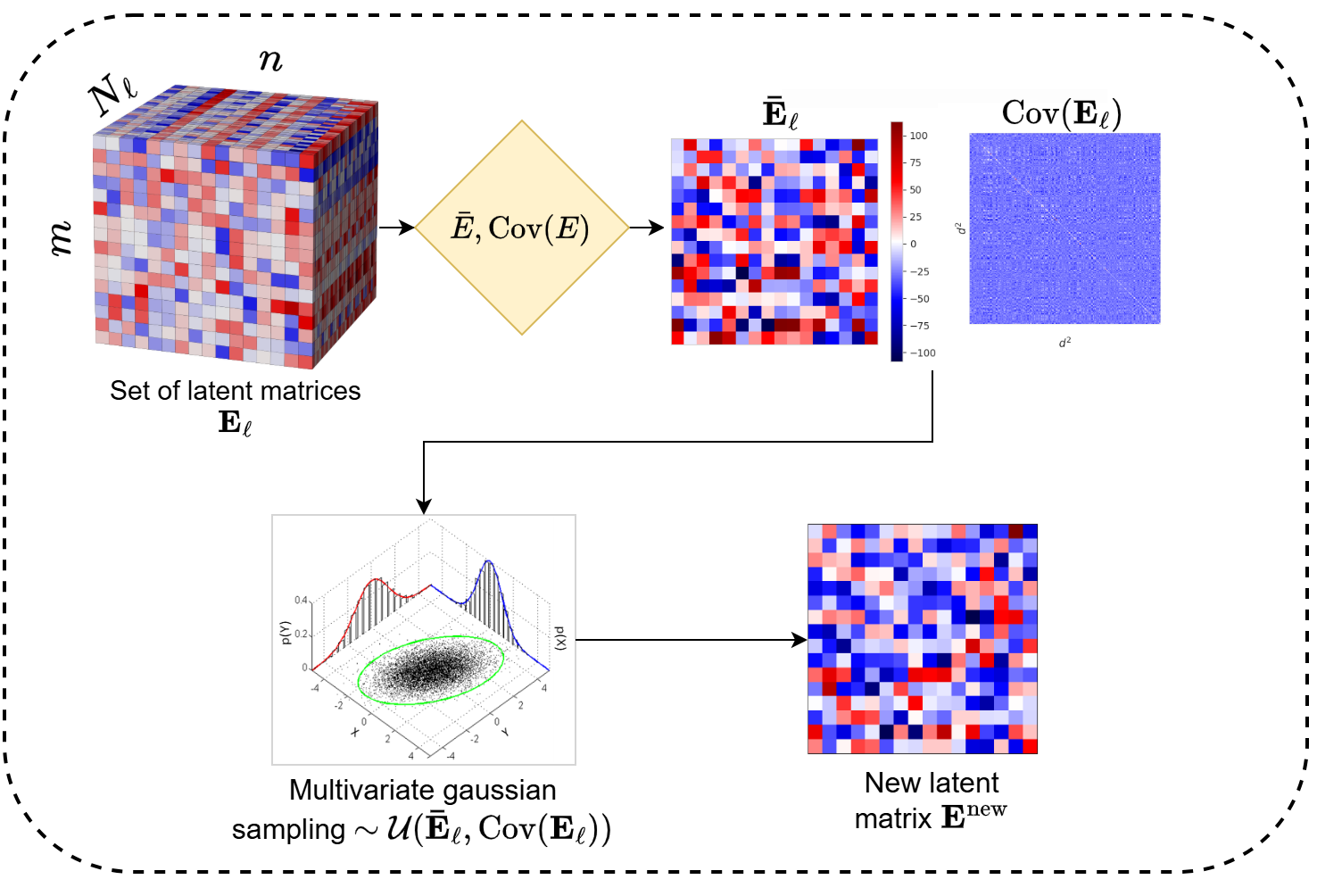}}
    \caption{Pipeline for generating synthetic latent matrices with a label \(\ell\) (label defining the chemical composition). Let \(\mathbf{\bar{E}_\ell}\) be the set of all latent matrices with label \(\ell\). The element-wise mean over all matrices \(\mathbf{\bar{E}_\ell}\) and covariance between each element \(\mathrm{Cov}(\mathbf{E}_\ell)\) are computed from \(\mathbf{E}_\ell\). New latent matrices are then sampled from the statistical distribution using multivariate Gaussian sampling around the mean \(\mathbf{\bar{E}_\ell}\) and the covariance \(\mathrm{Cov}(\mathbf{E}_\ell)\). To generate synthetic records, latent matrices are transformed back to the original space using the trained decoders \(g_\phi\) and \(g'_{\phi'}\).}
    \label{synthesization}
\end{figure}

From the latent matrices, new records can be synthesized by sampling new latent matrices from the statistical distribution of their respective group \(\mathbf{E}_\ell\), as is depicted in Fig \ref{synthesization}. First, the arithmetic average \(\mathbf{\bar{E}_\ell} \in \mathbb{R}^{d \times d}\) and the covariance between each matrix element \(\mathrm{Cov}(\mathbf{E}_\ell) \in \mathbb{R}^{d^2 \times d^2}\) are computed. New latent matrices are then generated by performing \textbf{multivariate Gaussian sampling} centred around \(\mathbf{\bar{E}_\ell}\) and with the computed covariance \(\mathrm{Cov}(\mathbf{E}_\ell)\):

\begin{equation}
    \mathbf{E}^{\text{new}} \sim \mathcal{U}(\mathbf{\bar{E}_\ell}, \mathrm{Cov}(\mathbf{E}_\ell)).
    \label{latent-generation}
\end{equation}

A newly synthesized record with features belonging to the specific group \(\ell\) can then be generated by sequentially applying both decoders \(g'_{\phi'}\) and \(g_\phi\) as described in \eqref{reconstruct-eq}.

\section{Experiments}

\subsection{Dataset}
\label{sec-dataset}

The pipeline has been evaluated on the \textit{HS-GC-IMS data of fermentations of different organisms} dataset \cite{bibdataset}. This dataset consists of 214 labelled GC-IMS records, representing both pure cultures and mixtures of the following organisms: \textit{E. coli} (EC), \textit{L. brevis} (LB), \textit{S. cerevisiae} (SC), and \textit{P. fluorescens} (PF). Mixtures always involve combinations of exclusively two organisms. There are in total $|\mathcal{L}| = 10$ different labels. The original GC-IMS records have dimensions of 6123 rows by 3150 columns.

\subsection{Preprocessing}
\label{preprocessing}

To improve data quality and processing efficiency, the Reactant Ion Peak line, which is redundant and present in all spectra, was removed. Given the large size and noisy nature of the original spectra, the resolution was reduced by wavelet compression \cite{bibwavelet} using the \textit{gc-ims-tools} toolbox \cite{bibgcimstools}. The wavelet transformation decomposes data into different frequency components, allowing efficient representation of the data. By discarding or compressing less significant coefficients, usually corresponding to high frequency noise or redundant details, significant storage savings can be achieved while retaining the core information. Additionally, a significant portion of the spectra was cropped, focusing on the region where peaks are present and discarding spaces in which no useful information is present.

The resulting 2D spectra have a reduced resolution of 769 rows by 174 columns. In this initial development of a record synthesization tool, a compressed version of the records is used to enable a more manageable computational load. However, the method is equally applicable to raw records.

\begin{figure}[htbp]
    \centerline{\includegraphics[width=0.5\textwidth]{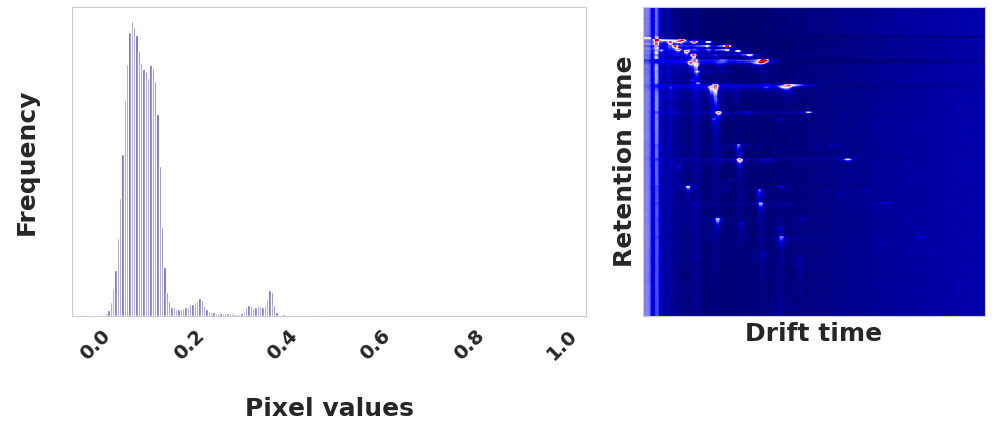}}
    \caption{Log-scaled record combined with min-max normalization (right) and the resulting histogram of pixel values (left). The log-scaling transforms the originally right-skewed histogram into a more balanced distribution, enhancing the detection of low-intensity peaks compared to the original records (see Fig. \ref{gcims}).}
    \label{logscaled}
\end{figure}

The analysis of the distribution of the data records shows a highly right-skewed distribution. Peaks, which are the most important features, only occupy a small fraction of the histogram. Log scaling combined with min-max normalization is applied to achieve a more balanced distribution, enhancing the visibility of low-intensity peaks otherwise overshadowed by dominant peaks (Fig. \ref{logscaled}). Finally, the method is reversible, allowing to rescale synthesized records back to the original distribution. 

\subsection{Architecture and Training}

The employed autoencoder architecture integrates a transformer encoder, fully connected layers positioned before and after the bottleneck, and a transformer decoder. Transformers are highly efficient sequential networks known for their state-of-the-art performance in tasks involving temporal or sequential data \cite{bibattention} \cite{bibtransformerae}. Their key strengths lie in the self-attention mechanism, which allows the model to capture temporal dependencies effectively, making them particularly suitable for analysing the sequential nature \cite{bibselfattention}, such as the shape of peaks and their position in GC-IMS records \cite{bibarchitectureae}.

The dataset was divided into an $85\%-15\%$ training-validation split. To ensure no \textit{data leakage}, the split was performed at the record level, and preprocessing steps were applied independently to the training and validation datasets, with the min-max normalization coefficients computed exclusively on the training data and applied on both sets.

Given that most timeseries (columns) in the GC-IMS records contain noise with no significant peaks, an undersampling strategy was implemented to balance the dataset. Specifically, the standard deviation of each timeseries was computed, and the median of these standard deviations was used as a threshold. Timeseries with a standard deviation below the median were assigned a $25\%$ chance of being included in the training dataset. This approach reduced the overrepresentation of "flat" or low-variance timeseries which exclusively contained noise, limiting a bias that could otherwise be transferred to the model during the training procedure. As a result, $19,684$ timeseries remained in the training dataset and $3,589$ in the validation dataset for the first autoencoder.

Training and validation losses were calculated using the MSE metric, as defined in \eqref{mse}. The learning rate was initialized at $0.0005$ and dynamically reduced when a plateau in the validation loss was detected. All models were trained for $150$ epochs using the Adam optimizer implementation by \textit{PyTorch} with default parameters \cite{bibpytorch}.

Because of the nature of the methodology (the number of rows are reduced $\mathbb{R}^m \rightarrow \mathbb{R}^d$ in the first autoencoder), the second autoencoder was trained with $1,810$ timeseries used for training and $330$ for validation. This training followed the same conditions, with the same learning rate and early stopping criteria. The training-validation split for this model is based on the same records as for the first autoencoder.

For both autoencoders, multiple latent vector sizes were explored during experimentation, specifically using the values $d = 8$, $d = 16$, and $d = 32$. The selection of this value represents a trade-off between computational efficiency for the training and synthetisation steps and the reconstruction quality of the autoencoders. A larger value for $d$ allows the model to project the timeseries into a larger latent space, thereby retaining more information when reconstructing back to the original space. However, larger values of $d$ exponentially increases the size of the covariance matrix, which significantly slows down the multivariate Gaussian sampling algorithm.

\subsection{Synthetisation}

The GC-IMS records were grouped by their label $\ell \in \mathcal{L}$, where $\mathcal{L}$ is described in section \ref{sec-dataset}. For every number of matrices in a group $N_\ell$, the same number of spectra were generated, effectively doubling the number of GC-IMS records.

\subsection{Classification}

% To evaluate the quality of the generated records, a \textbf{discriminator model} \cite{bibgan} was fitted on a training set containing an equal representation of both original and synthetic records, to avoid bias. After training, the model's performance was evaluated on a validation set to assess the quality of the generated records. This performance is quantified using an accuracy rating (AR), which is computed as the ratio of correctly classified records as "original" or "synthetic." An AR close to $50\%$ suggests that the model is unable to distinguish between "original" and "synthetic" records, which typically indicates good performance. It is crucial that the AR computed over the training dataset is similar to that computed over the validation dataset. If the AR on the training set is significantly higher, it would suggest that the model has overfitted to the dataset.

% As the discriminator, a pipeline was trained on a dataset containing an equal number of original and synthetic records. The records were log-normalized as described in Section \ref{preprocessing}, then processed with Principal Component Analysis (PCA) to reduce their dimensionality from 769 rows and 174 columns to only 22 components. Finally, the records were classified from the components as either "original" or "synthetic" using a Random Forest Classifier, configured with the default settings from \textit{Scikit-learn's} implementation \cite{bibsklearn}. To ensure the statistical robustness of the results, a total of $5$ iterations were trained using a $85\%-15\%$ training-validation split.

The objective to develop a classification model is to assess whether augmenting the training dataset with synthetic records improves classification performance. A \textbf{record classifier} model was therefore trained. Such model predicts the label $\ell \in \mathcal{L}$ of a given GC-IMS record. Demonstrating a measurable improvement in classification accuracy when synthetic records are included would validate the utility of generating synthetic records.

For the classifier, records were preprocessed as described in Section \ref{preprocessing}. Their dimensionality was reduced from 769 rows and 174 columns to 22 components with a Principal Component Analysis (PCA). This number of components was chosen to achieve an explained accumulated variance of over $90.00\%$. The records were subsequentially classified using a random forest model configured with the default settings from the \textit{Scikit-Learn's} implementation \cite{bibsklearn}. To ensure the statistical robustness of the results, a total of $5$ models were trained using different training and validation splits. For each model, performance was measured using the Accuracy Rating (AR) score, which is the ratio of correctly classified records.

\section{Results}

\subsection{Autoencoders}

Table \ref{autoencoder-training} shows that increasing the size of the latent vector $d$ improved the reconstruction quality of the signals by the autoencoders, as evidenced by the MSE \eqref{mse} over the validation dataset. This improvement occurred because a larger latent vector provided the autoencoder with a higher capacity to encode and preserved more information from the original signal. However, this effect exhibited diminishing returns, making it important to choose an optimal size for the latent vector. Additionally, increasing $d$ resulted in greater computational requirements for the generation of synthetic records (Section \ref{sec-synthetic}).

% Fig. \ref{reconstruction-ret} illustrates how the first autoencoder reconstructs signals along the retention axis. The "pulses" in the timeseries correspond to the peaks in the GC-IMS records, making their accurate reconstruction in terms of position and amplitude essential. The second-largest peak in the figure represents a less frequent occurrence in the dataset; consequently, models with lower $d$ tended to under-reconstruct its amplitude, as they prioritized capturing the most common features with their limited capacity.

% Fig. \ref{reconstruction-drift} illustrates the reconstruction of by the second autoencoder. Since each model was traineded on data derived from a separate model with the same latent dimension $d$, it was not possible to generate a figure with three reconstructions of the same timeseries. For example, the first figure with $d = 8$ was trained from the latent vectors obtained by the first autoencoder with $d = 8$, while the second figure with $d = 16$ was trained over the latent vectors from the first autoencoder with $d = 16$. Regardless, the conclusions of the impact of $d$ on reconstruction quality were consistent with those observed for the first autoencoder.

\begin{figure}[htbp]
    \centerline{\includegraphics[width=0.5\textwidth]{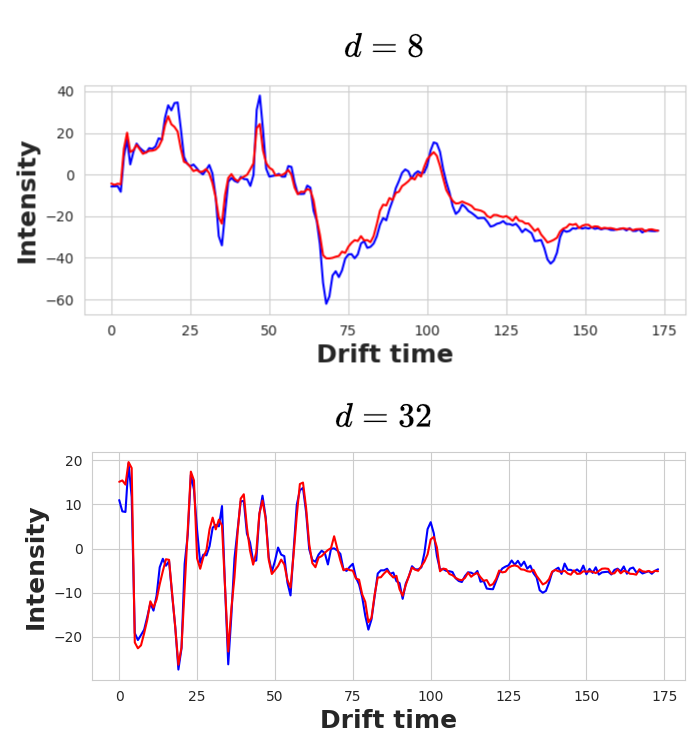}}
    \caption{Reconstruction by the second autoencoder (in red) of the evolution along the drift axis of latent vectors (in blue). Using the transformation: $\hat{\mathbf{x}} = (g'_{\phi'} \circ f'_{\theta'} \circ f_\theta)(\mathbf{x})$. Each line corresponds to a different model, as the input data were processed separately for each latent dimension $d$. Increasing $d$ enhances reconstruction quality.}
    \label{reconstruction-drift}
\end{figure}

\begin{figure}[htbp]
    \centerline{\includegraphics[width=0.5\textwidth]{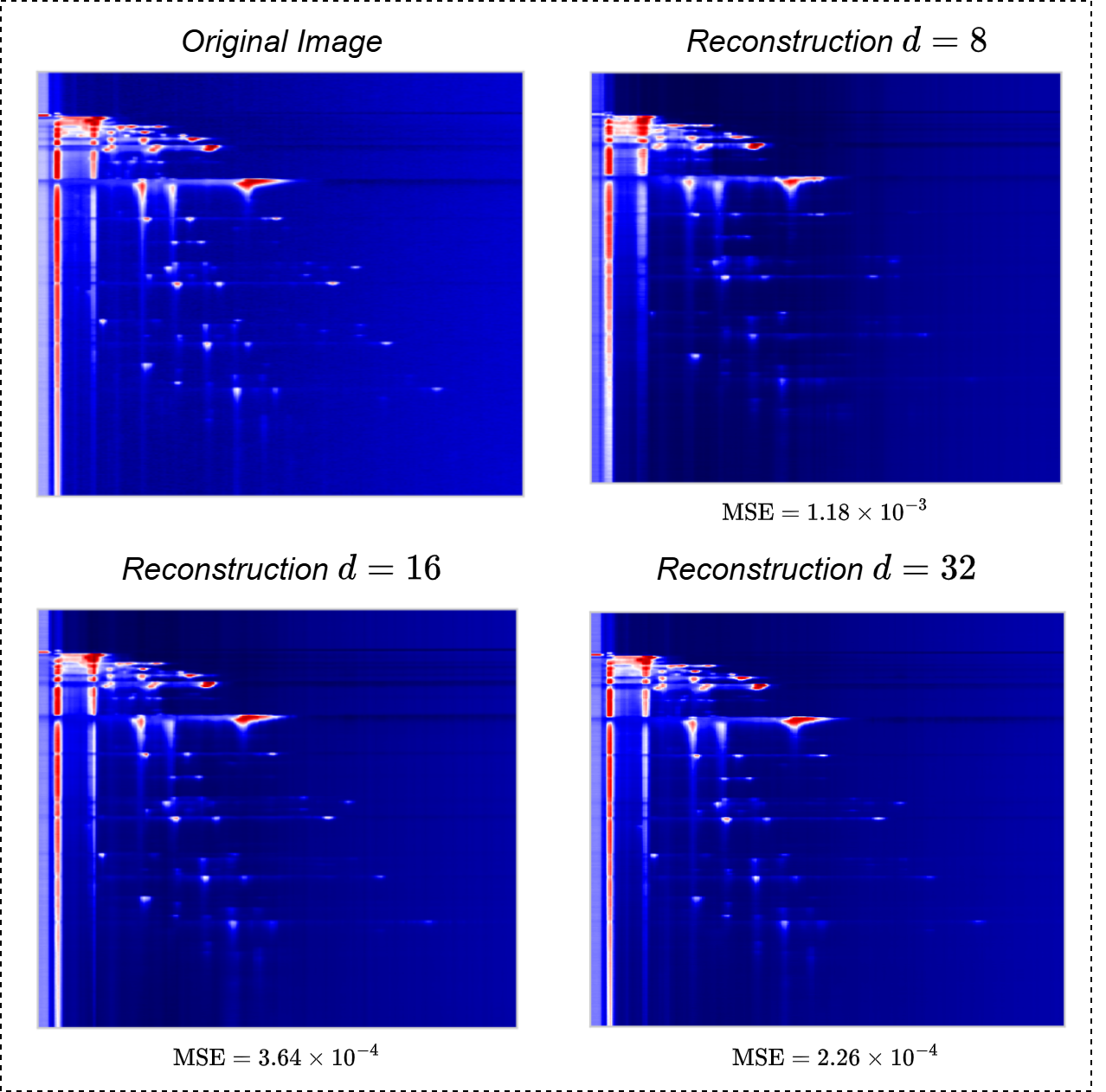}}
    \caption{Reconstruction of records from the mixed class \textit{E. coli and S. cerevisiae} using autoencoders with latent vector dimensions of $d = 8$, $d = 16$, and $d = 32$, respectively. Increasing the latent vector size $d$ enhances the reconstruction of less prominent features, such as smaller peaks concentrated in the lower regions of the records. Additionally, the shapes of dominant peaks are more accurately reconstructed as $d$ increases.}
    \label{reconstruction-records}
\end{figure}

Fig. \ref{reconstruction-drift} displays the reconstruction of timeseries by the second autoencoder, while Fig. \ref{reconstruction-records} illustrates the reconstruction of a complete record obtained by combining both decoders: $(g_\phi \circ g'_{\phi'} \circ f'_{\theta'} \circ f_\theta)(\mathbf{x})$. Records reconstructed using models with smaller latent dimensions $d$ tend to lose finer details, particularly the smaller peaks located in the lower regions of the records. As reflected in the MSE for each reconstruction, increasing the latent dimension improved reconstruction quality although exhibited diminishing returns.

For the subsequent experiments, the latent vector size were fixed at $d = 32$ for both autoencoders, as further increasing $d$ would yield limited benefits while significantly increasing computational demands.

\begin{table}[htbp]
    \caption{MSE and Training time for autoencoders}
    \begin{center}
    \begin{tabular}{|c|c|c|c|c|c|c|}
    \hline
    \textbf{Latent}&\multicolumn{2}{|c|}{\textbf{Mean Squared Error}}&\multicolumn{2}{|c|}{\textbf{Training time}} \\
    \cline{2-5} 
    \textbf{vector $d$} & \textbf{\textit{Retention}}& \textbf{\textit{Drift}}& \textbf{\textit{Retention}}& \textbf{\textit{Drift}} \\
    \hline
    $d = 8$ & $4.35 \times 10^{-4}$ & $5.26 \times 10^{1}$ & $1.705$ hours & $1.596$ mins \\
    \hline
    $d = 16$ & $2.78 \times 10^{-4}$ & $9.61 \times 10^{0}$ & $1.754$ hours & $3.604$ mins \\
    \hline
    $d = 32$ & $1.94 \times 10^{-4}$ & $6.62 \times 10^{0}$ & $1.761$ hours & $7.123$ mins \\
    \hline
    \end{tabular}
    \label{autoencoder-training}
    \end{center}
\end{table}

% \begin{figure}[htbp]
%     \centerline{\includegraphics[width=0.5\textwidth]{figures/ts_reconstruction_ret.drawio.png}}
%     \caption{Reconstruction (in red) of the original signals along the retention axis (in blue) by the first autoencoder using the transformation: $ \hat{\mathbf{x}} = (g_\phi \circ f_\theta)(\mathbf{x})$. Each panel presents the same line from a single record, reconstructed using models with latent vectors of different dimensions. As the latent dimension increases, the autoencoder reconstructs with more information, resulting in higher reconstruction quality, albeit with diminishing returns.}
%     \label{reconstruction-ret}
% \end{figure}

% \begin{figure}[htbp]
%     \centerline{\includegraphics[width=0.5\textwidth]{figures/ts_reconstruction_drift.drawio.png}}
%     \caption{Reconstruction (in red) of the evolution along the drift axis of latent vectors computed by the first autoencoder (in blue). Using the transformation: $\hat{\mathbf{x}} = (g'_{\phi'} \circ f'_{\theta'} \circ f_\theta)(\mathbf{x})$. Each line corresponds to a different model, as the input data were processed separately for each latent dimension $d$. As before, increasing $d$ enhances reconstruction quality, with diminishing returns.}
%     \label{reconstruction-drift}
% \end{figure}

\subsection{Synthetic Records}
\label{sec-synthetic}

\begin{figure}[htbp]
    \centerline{\includegraphics[width=0.5\textwidth]{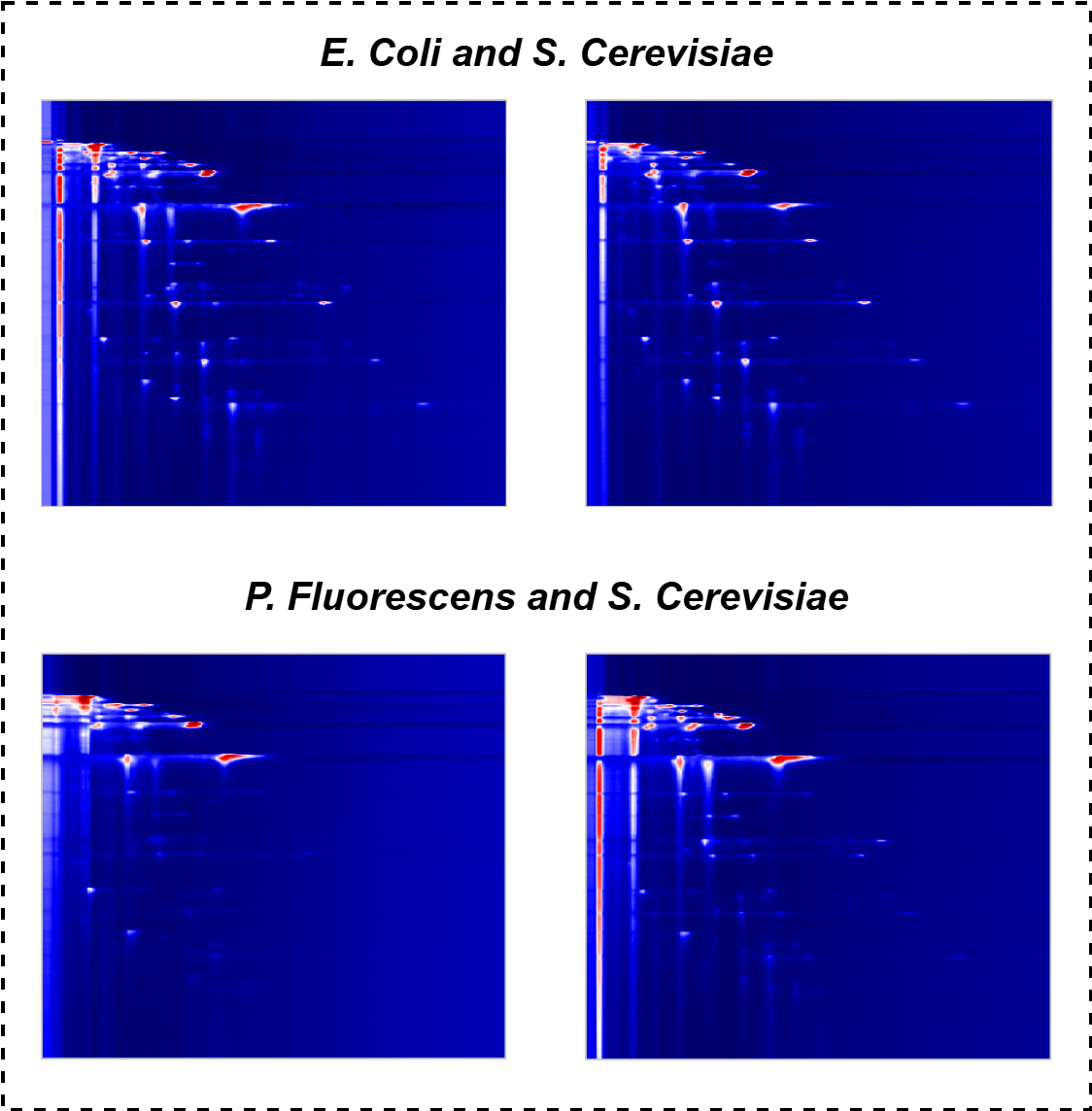}}
    \caption{Synthetic records for the mixed classes \textit{E. coli and S. cerevisiae} and \textit{P. fluorescens and S. cerevisiae}. The variability in the synthesized records is evident in the intensity and shape of the peaks, as well as a slight shift in the overall signature within the GC-IMS spectrum. This variability is consistent with that observed in the original dataset.}
    \label{synthesization-records}
\end{figure}

Fig. \ref{synthesization-records} illustrates synthetic records for the mixed classes \textit{E. coli and S. cerevisiae} and \textit{P. fluorescens and S. cerevisiae}. Similar to reconstructed records, features are lost due to the inherent nature of autoencoders. However, the synthetic records provide new samples with dominant features that closely resemble those present in the original distribution. Additionally, the variability in the synthetic records realistically reflects the variability observed in the original distribution.

\subsection{Classification}

\begin{figure}[htbp]
    \centerline{\includegraphics[width=0.5\textwidth]{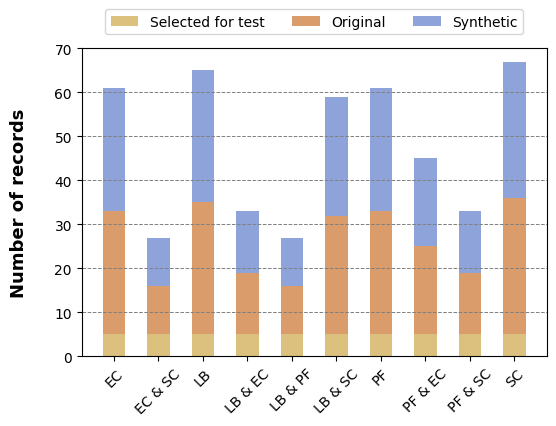}}
    \caption{Number of samples per label ($\ell$). The dataset size is increased by doubling the records with synthetic samples, addressing the scarcity of data in mixed classes such as \textit{E. coli and S. cerevisiae} ("EC \& SC"). During training, test data were exclusively drawn from the original samples, as synthetic data were not suitable for model evaluation.}
    \label{class-representation}
\end{figure}

The class representation of the dataset is displayed in Fig. \ref{class-representation}. To address the issue of sample scarcity in mixed classes, such as \textit{E. coli and S. cerevisiae}, the number of records was augmented by a factor of two by introducing synthetic records into the dataset while maintaining the same class proportions as in the original dataset. Since synthetic records cannot be used to validate the performance of the model, the validation samples were drawn exclusively from the original records, with a quantity of 5 samples per class.

% Table \ref{class-representation} illustrates the performance in terms of Accuracy Rating (AR) for both the discriminator and record classifier models. The discriminator demonstrated a high proficiency in differentiating between original and synthetic records. Maintaining the multiplicative factor $\alpha$ close to $1$ resulted in the generation of samples that closely resemble the original distribution, thereby diminishing the distinguishable features between original and synthetic samples.

% Furthermore, the discriminator exhibited reduced performance when attempting to differentiate between \textbf{recreated samples}—samples processed through the autoencoder without alterations in the latent matrix—and synthesized samples. This phenomenon can be attributed to the inherent information loss induced by autoencoders, particularly for lower-dimensional latent spaces ($d$). The loss of detail in synthesized records made the comparison between recreated and synthesized samples a more challenging task for the model, as opposed to differentiating original samples from synthesized ones.

% Regardless, the performance of the discriminator when differentiating original and synthetic samples is the most important metric of quality among the two. The discriminator's accuracy of $67.44\%$ indicates that synthesized samples are still quite distinguishable from the original ones.

Augmenting the dataset with synthetic records resulted in a \textbf{significant improvement} in the AR, increasing from an average of $75.60\% \pm 3.44$ using only original records as a baseline, to $84.40\% \pm 3.21$ ($\Delta_{AR} \simeq 8.9\%$) with the augmented dataset. Additionally, models were trained using a dataset that combined reconstructed records (records encoded and decoded by the autoencoders without alterations to their latent matrices) and synthetic records. These models yielded slightly inferior results ($\mathrm{AR} = 83.60\% \pm 2.71$) compared to the original dataset augmented with synthetic records.

% \begin{table*}[htbp]
%     \caption{Data for different microorganism combinations}
%     \begin{center}
%     \begin{tabular}{|c|c|c|c|c|c|c|c|c|c|c|}
%     \hline
%     \textbf{Dataset} & \textbf{LB} & \textbf{EC} & \textbf{PF} & \textbf{SC} & \textbf{LB \& EC} & \textbf{LB \& PF} & \textbf{LB \& SC} & \textbf{PF \& EC} & \textbf{EC \& SC} & \textbf{PF \& SC} \\
%     \hline
%     \textbf{Original - Training} & 25 & 23 & 23 & 26 & 9 & 6 & 22 & 15 & 6 & 9 \\
%     \hline
%     \textbf{Original - Validation} & 5 & 5 & 5 & 5 & 5 & 5 & 5 & 5 & 5 & 5 \\
%     \hline
%     \textbf{Synthetic - Training} & 30 & 28 & 28 & 31 & 14 & 11 & 27 & 20 & 11 & 14 \\
%     \hline
%     \end{tabular}
%     \label{class-representation}
%     \end{center}
% \end{table*}

% \begin{table*}[htbp]
%     \caption{Accuracy scores for the discriminator and classifier models.}
%     \begin{center}
%     \begin{tabular}{|c|c|c|c|c|c|}
%     \hline
%     \multirow{2}{*}{\textbf{$\alpha$}}&\multicolumn{2}{|c|}{\textbf{Discriminator accuracy}}&\multicolumn{3}{|c|}{\textbf{Classification accuracy}} \\
%     \cline{2-6}
%      & 
%     \fontsize{7}{10}\selectfont \textit{\textbf{Original vs. Synthetic (\%)}} & 
%     \fontsize{7}{10}\selectfont \textit{\textbf{Recreated vs. Synthetic (\%)}} & 
%     \fontsize{7}{10}\selectfont \textit{\textbf{Baseline (Original only) (\%)}} & 
%     \fontsize{7}{10}\selectfont \textit{\textbf{Original + Synthetic (\%)}} & 
%     \fontsize{7}{10}\selectfont \textit{\textbf{Recreated + Synthetic (\%)}} \\
%     \hline
%     $0.1$ & $95.66 \pm 2.48$ & $97.67 \pm 0.69$ & $75.60 \pm 3.44$ & $76.00 \pm 2.19$ & $74.80 \pm 0.98$ \\
%     $0.5$ & $89.15 \pm 2.77$ & $86.20 \pm 2.10$ & $75.60 \pm 3.44$ & $78.40 \pm 2.94$ & $77.60 \pm 2.94$ \\
%     $1$ & $67.44 \pm 3.89$ & $\mathbf{61.86 \pm 5.83}$ & $75.60 \pm 3.44$ & $\mathbf{84.40 \pm 3.21}$ & $83.60 \pm 2.71$ \\
%     $2.5$ & $78.91 \pm 2.37$ & $82.17 \pm 3.21$ & $75.60 \pm 3.44$ & $81.20 \pm 2.40$ & $79.20 \pm 2.81$ \\
%     $5$ & $91.47 \pm 2.35$ & $92.71 \pm 1.81$ & $75.60 \pm 3.44$ & $75.60 \pm 4.08$ & $76.00 \pm 2.83$ \\
%     $10$ & $98.29 \pm 1.66$ & $98.29 \pm 1.73$ & $75.60 \pm 3.44$ & $70.00 \pm 2.83$ & $72.40 \pm 3.44$ \\
%     \hline
%     \end{tabular}
%     \label{classification-perfomance}
%     \end{center}
% \end{table*}

\section{Conclusion}

In this publication, a general method for synthesizing 2D data records from a limited dataset was introduced. By exploiting the correlations present within the rows and columns of the records, a double autoencoder architecture was developed to compress these records into compact latent matrices. Subsequently, new latent matrices are generated modifying the original latent matrices based on their statistical distribution across label classes. Ultimately, synthetic records in the original data space are generated by decoding these new generated latent matrices using the trained decoders.

Furthermore, experiments were conducted on a publicly available dataset of GC-IMS spectra, demonstrating the impact of the autoencoders' latent dimension parameter $d$. It was then demonstrated that expanding a training dataset by including synthetic records significantly enhances the performance of a label classification pipeline ($+8.9\%$ on the classification accuracy) by introducing greater variety. This, in turn, improves model robustness and addresses the scarcity of records for underrepresented labels.

Future research should focus on mitigating the information loss inherent in the autoencoder, as it has been demonstrated that this loss manifests itself in the synthetic samples, which often fail to capture finer details. Additionally, further applications in the synthesization of 2D data spectra could be explored, such as Fluorescence Spectroscopy in 2D maps, Microarray Analysis, or any 2D frequency spectra of timeseries signals such as electrocardiography or electroencefalography.

% \section*{Acknowledgment}

% This project was funded by the European Union through Grant Agreement 101103176. The views and opinions expressed are those of the author(s) only and do not necessarily reflect those of the European Union or the European Commission. Neither the European Union nor the granting authority can be held responsible for them. 

\end{document}